\documentclass[aps,prc,twocolumn,floatfix,showpacs,amsmath,amssymb,nofootinbib]{revtex4-1}
\usepackage[T1]{fontenc}
\usepackage{graphicx}
\usepackage{bm}
\usepackage{color}
\usepackage{dcolumn}

\begin{document}

\title{Nuclear Pasta Matter for Different Proton Fractions}

\author{B. Schuetrumpf$^1$, K. Iida$^3$, J. A. Maruhn$^1$, P.-G. Reinhard$^2$}

\affiliation{
$^1$Institut f\"ur Theoretische Physik, Universit\"at Frankfurt, D-60438 Frankfurt, Germany 
} 
\affiliation{
$^2$Institut f\"ur Theoretische Physik, Friedrich-Alexander-Universit\"at Erlangen-N\"urnberg, D-91058 Erlangen, Germany
} 
\affiliation{
$^3$Department of Natural Science, Kochi University, 2-5-1 Akebono-cho, Kochi 780-8520, Japan
} 

\date{\today}

\begin{abstract}
Nuclear matter under astrophysical conditions is explored with time-dependent
and static Hartree-Fock calculations.  The focus is in a regime of densities
where matter segregates into liquid and gaseous phases unfolding a rich scenario
of geometries, often called nuclear pasta shapes (e.g. spaghetti, lasagna). 
Particularly the appearance of the different phases depending on the proton
fraction and the transition to uniform matter are investigated. In this context
the neutron background density is of special interest, because it plays a
crucial role for the type of pasta shape which is built. The study is performed
in two dynamical ranges, once for hot matter and once at temperature zero to
investigate the effect of cooling.
\end{abstract}


\maketitle

\section{Introduction}

Nuclear matter in supernova cores and proto-neutron stars covers a wide range of 
density, temperature, and proton fraction, and hence exhibits various 
interesting properties such as, e.g., liquid-gas mixing \cite{Lamb}.  It is 
believed that after the bounce of the inner core, cooling and deleptonization 
lead to drastic changes in the temperature and proton fraction of each matter 
element inside the core \cite{Bethe,Suzuki}. Just below normal nuclear density, 
matter is likely to disentangle into exotic shapes, which resemble pastas like 
spaghetti and lasagna and can affect neutrino transport 
\cite{Ravenhall,Hashimoto,Lassaut,Pethick,Chamel,Horowitz2004,Horowitz20042, 
Sonoda2007}. Steadily increasing computing power allows meanwhile to simulate 
pasta matter by fully three-dimensional, symmetry unrestricted time-dependent 
Hartree-Fock (TDHF) calculations. This has led to a revival of investigations of 
structure and dynamics of pasta matter within TDHF 
\cite{NewtonStone,Sebille,Sebille2011,Pais,Schuetrumpf2013a} as well as 
classical and quantum molecular dynamics calculations 
\cite{Sonoda2008,Schneider2013}. In this contribution, we address the question 
how the map of pasta shapes changes with the proton fraction and how the neutron 
excess is distributed between the nuclear liquid and a gaseous neutron 
background.  Nuclear pasta matter for different values of the proton fraction 
has been studied so far using classical molecular dynamics \cite{Dorso}, the 
TDHF-like DYWAN model\cite{Sebille2011}, and a Thomas-Fermi approximation based 
on a relativistic mean-field model \cite{Okamoto}. Here we aim at a fully 
quantum mechanical TDHF description. We thus extend our previous TDHF 
calculations for proton fraction 1/3 \cite{Schuetrumpf2013a} to the general case 
of arbitrary proton fractions. Thereby we concentrate on the regime of low 
densities up to saturation density of symmetric nuclear matter. We consider 
phases at high temperature in the MeV range which are relevant for core-collapse 
super-novae as well as phases at zero temperature which have relevance to 
materials that deleptonize in proto-neutron stars.

For the calculations done here the Skyrme-TDHF code as explained in 
\cite{Maruhn2014} is used. For the astrophysical calculations periodic boundary 
conditions are assumed. For the interaction the Skyrme force SLy6 is taken 
\cite{Chabanat}. Additionally to the proton charge distribution we take a 
uniform electron background to keep the charge neutrality. 

\section{Simulation and basic quantities}

Static and dynamic properties of inhomogeneous nuclear matter are computed with 
the TDHF code. For the astrophysical calculations performed here, periodic 
boundary conditions are applied, thus simulating infinite matter approximately 
with a periodic lattice of simulation boxes. The imposed periodicity exerts a 
constraint on the system which, in turn, may shift a bit the transition points 
from one phase to another. We have checked the impact of box size earlier 
\cite{Schuetrumpf2013a} and find that effects are not dramatic for the simple 
geometries discussed here, although box size can become an issue when dealing 
with involved shapes as, e.g., gyroids \cite{Schuetrumpf2014a}. The actual 
simulation box uses a grid spacing of 1 fm and 16$\times$16$\times$16 grid 
points thus having a box length $l=16$ fm and a box volume $V=l^3$.  We vary the 
number of $\alpha$-particles $N_\alpha$, and of additional neutrons $N_{n+}$. 
The Coulomb field in a periodic simulation is given by the Ewald sum 
\cite{Ewa21a}. The Fourier representation of kinetic energy allows a very 
elegant computation in Fourier space. Total charge neutrality is achieved by 
assuming a homogeneous electron background. This approximation ignores screening 
effects from the electrons, for detailed discussions see 
\cite{Mar05b,Watanabe2003a,Dorso2012,Alcain}.  Although screening lengths can 
vary in a wide range and are occasionally of the order of structure size, the 
net effect of electron screening is found to be small 
\cite{Mar05b,Watanabe2003a}, thus justifying the homogeneous approximation for 
the present exploration.

The initial state is prepared in similar fashion as in \cite{Schuetrumpf2013a}. 
From the $N_p$ protons together with $N_p$ neutrons, we form $N_\alpha=N_p/2$ 
$\alpha$-particles. These are distributed stochastically over the whole space of 
the simulation box while keeping a minimal distance of ${3.5\rm\,fm}$ between 
the centers of the $\alpha$-particles to avoid too large overlaps between them. 
The $\alpha$-particles are initialized at rest. The remaining $N_n-N_p$ neutrons 
are initialized as plane waves filling successively the states with the lowest 
kinetic energies with degenerate pairs of spin-up and spin-down particles. 
Finally, all nucleons are ortho-normalized. The then following time evolution of 
the system is done by standard TDHF propagation \cite{Maruhn2014}. 

This initial state is far above the ground state. During the first few fm/c of 
dynamical evolution, the system quickly evolves into a fluctuating thermal state 
whose average properties (shapes, densities, energies) remain basically 
constant.  The emerging temperatures end up in a range well representing pasta 
matter in core-collapse supernovae. They lie between 2 and 7 MeV depending on 
proton fraction $X_p$ and density $\rho$ where lower $X_p$ and larger $\rho$ are 
found to be associated with lower temperatures. For the present exploratory 
stage, we use just this broad temperature range as representatives of hot 
matter. In a second step, we cool down the system for each $X_p$ and density 
$\rho$ to a locally stable zero-temperature state. To that end, we start from a 
given dynamical state and perform a static calculation with standard techniques 
of the code \cite{Maruhn2014}. The starting configuration is taken from some 
time point in the late phases of the TDHF simulation. The precise time is 
unimportant for the final result. The whole procedure (random initialization, 
evaluation of thermal state, cooling) is done twice for each setting of $X_p$ 
and $\rho$. This gives some clue on the stability of a configuration.

A word is in order about the analysis of non-homogeneous structures of infinite 
systems in a finite simulation box. It is known that the imposed boundary 
conditions have an impact on the structures due to symmetry violation and 
possibly spatial mismatch \cite{Hoc81aB,All87}. A study within classical 
molecular dynamics without Coulomb interactions implies that this may be 
particularly critical for extended, inhomogeneous structures \cite{Gimenez}. 
However, the numerical expense of fully quantum-dynamical simulations sets 
limits on the affordable box sizes. To check its effect, we have studied for the 
case of $X_p=1/3$ also larger boxes up to 26 fm and find the same structures. 
Thus we confine the exploratory survey here to the one box size 16 fm.

In order to characterize the matter, we will use a few global properties. The
total number of protons and neutrons is
\begin{equation}
  N_p=2N_\alpha
  \quad,\quad
  N_n=2N_\alpha+N_{n+}
\end{equation}
where $N_\alpha$ is the number of initial $\alpha$ particles (strictly related
to $N_p$) and $N_{n+}$ the number of excess neutrons (not absorbed into initial
$\alpha$ particles). The key parameter is the proton fraction $X_p$
\begin{equation}
  X_p=\frac{N_p}{N_p+N_n}=\frac{2N_\alpha}{2N_\alpha+N_{n+}}\leq\frac{1}{2}
  \quad.
\end{equation}
The latter inequality means that we confine studies to neutron rich systems
which is the typical scenario in astrophysical environment. There are several
further measures related to densities.  The most important one is the mean
density
\begin{equation}
  \rho
  =
  \frac{N_p+N_n}{V}
  =
  \frac{4N_\alpha+N_{n+}}{V}
  =
  \frac{A}{V}
\end{equation}
where $V$ is the box volume and $A=N_p+N_n$ is the total nucleon number. It is 
only at high densities that matter stays homogeneous. Usually, matter segregates 
into dense regions of nuclear liquid with density $\rho_l$ and a dilute gas 
phase with density $\rho_g$, consisting here of a neutron gas.  These two phases 
fill correspondingly volumes $V_l$ and $V_g$ with $V_l+V_g=V$. These volumes are 
defined with the help of the Gibbs dividing surface \cite{Shchukin}. A threshold 
density $\rho_\mathrm{thr}$ is set and all regions with 
$\rho(\mathbf{r})>\rho_\mathrm{thr}$ are added to $V_l$ and all other to $V_g$. 
The threshold is determined such that
\begin{equation}
  \rho_l\cdot V_l+\rho_g\cdot V_g
  =
  A
  \quad.
\end{equation}
From these volumes, the volume fraction $u_l$ of the liquid phase can be deduced 
as
\begin{equation}
  u_l
  =
  \frac{V_l}{V}
  =
  \frac{\rho-\rho_g}{\rho_l-\rho_g}
  \quad.
\label{eq:Gibbs}
\end{equation}
Since $\rho$ is always smaller than $\rho_l$ for pasta shapes and the gas 
density consists of the background densities, the volume fraction becomes 
smaller for larger neutron background densities and the type of pasta shape 
which is formed depends strongly on this volume fraction.

The volume fraction $u_l$ is the most important parameter to determine the 
geometry of the system. With steadily increasing $u_l$, the system marches 
through a series of geometrical phases.  For the lowest $u_l$ the individual 
shapes are the sphere, which corresponds to finite nuclei with large zones of 
vacuum around.  In the next stage, the nuclei fuse to cylinder-like shapes, the 
``rod'' structure, also denoted as ``spaghetti''. Further increasing $u_l$ leads 
to planar meshes of orthogonal rods, ``rod(2)'', followed by three-dimensional 
grids of rods, denoted ``rod(3)''. Equally densely packed is the ``slab'' 
structure, parallel planes completely filled with matter, also called in a more 
appetizing manner ``lasagna''. From then on the picture is reversed. We have 
more matter than voids. The slab and rod(3) are symmetric under exchange of 
matter and voids, thus residing at the turning point. Further increasing $u_l$ 
then yields the phases of ``rod(2) bubble'', ``rod bubble'', and ``sphere 
bubble'', where bubble denotes that the gas phase has the shape of the pasta. A 
further increase ends up in homogeneous matter. These are the phases which we 
will study in the following.

\section{Results}
\begin{figure}[tb]
\includegraphics[width=\linewidth]{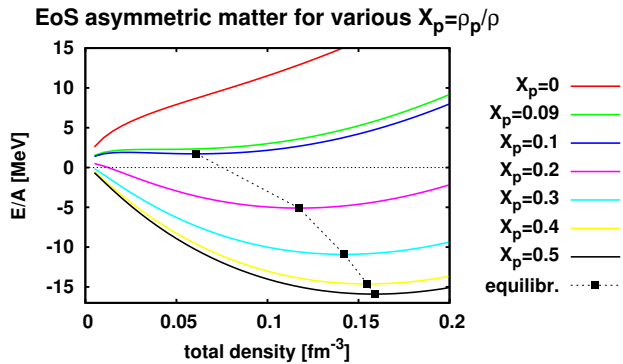}%
\caption{\label{fig:EoS-protfrac}(Color online) Binding energy per
  particle $E/A$ as function of total density for asymmetric nuclear
  matter at zero temperature for various proton fractions $X_p$ as
  indicated. The ground states are indicated by black boxes for those
  $X_p$ where a local minimum could be found. The curves have been
  computed for the Skyrme force SLy6.}
\end{figure}

\begin{figure}[tb]
\includegraphics[width=0.8\linewidth]{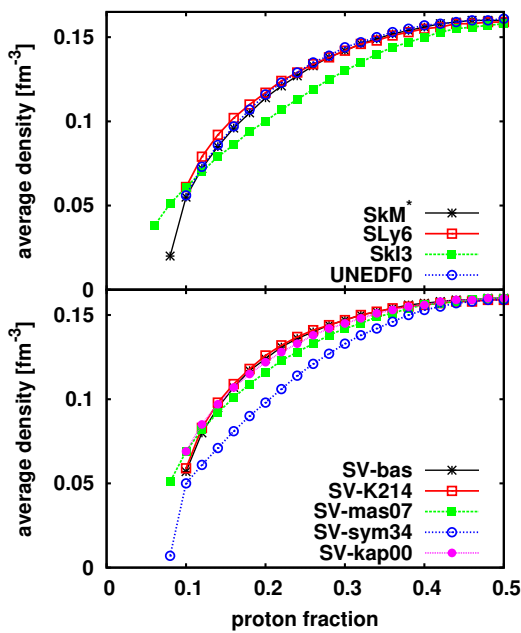}%
\caption{\label{fig:vary-asymnucmat}(Color online) Equilibrium density
  as function of proton fraction $X_p$ for different Skyrme
  interactions (explanations see text). Each curve stops at some low
  $X_p$ because no equilibrium (minimum of energy as function of
  $\rho$) could be found for smaller $X_p$.  }
\end{figure}

\begin{figure*}[t!]
%\newsavebox\XPBox
%\savebox\XPBox{
\includegraphics[width=0.95\linewidth]{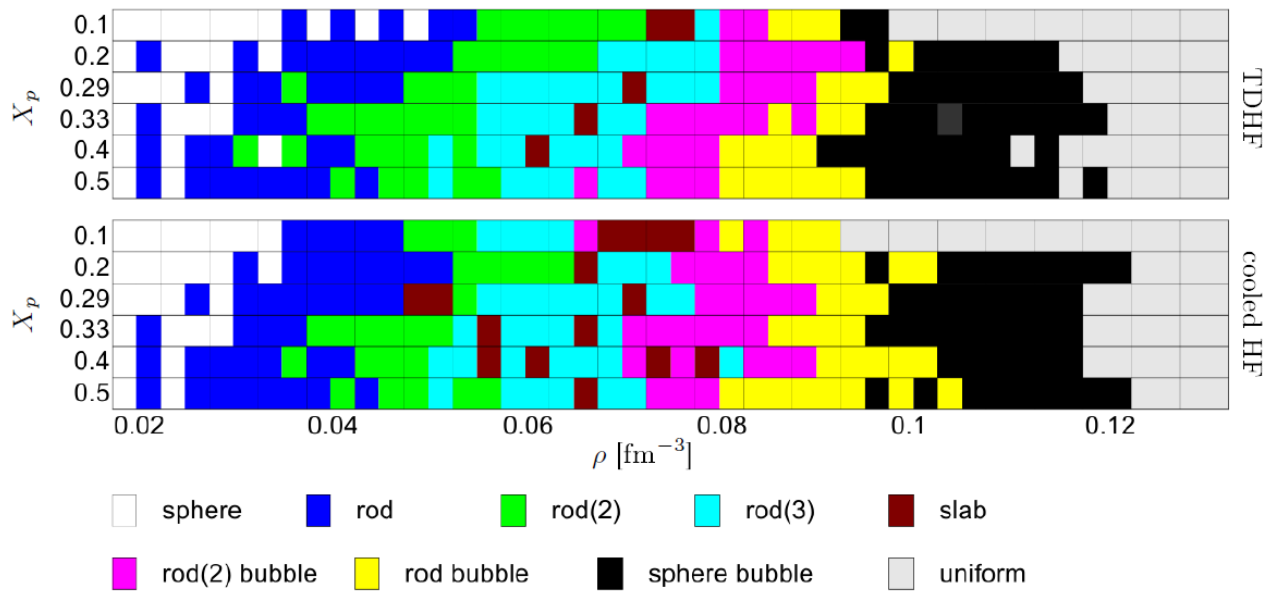}
%}
%  \begin{pspicture}(0,0)(\wd\XPBox,\ht\XPBox)
%     \rput[lb](0.,0.){\usebox\XPBox}
%     \rput[c]{90}(0.45,3.28){$X_p$}
%     \rput[c]{90}(0.45,5.65){$X_p$}
%     \rput[rb](8.1,1.5){$\rho$ [fm$^{-3}$]}
%     \rput[c]{270}(14.5,3.28){cooled HF}
%     \rput[c]{270}(14.5,5.65){TDHF}
%   \end{pspicture}
\caption{\label{fig:phadiag_X_P}(Color online) Upper panel: Map of pasta shapes 
achieved in TDHF calculations starting from a gas of $\alpha$-particles with 
neutron background for various proton fractions and mean densities. Starting 
from these final states, the map of pasta shapes achieved in Hartree-Fock 
calculations (cooled pasta) is shown in the lower panel. Two different colors in 
a cell mark different final states for two calculations with similar initial 
conditions. The structures are named like in \cite{Schuetrumpf2013a} and 
examples and further properties can be found there.}
\end{figure*}

\subsection{Equation of state of homogeneous matter}

As a first step, we look at the binding energy curves of homogeneous asymmetric 
nuclear matter for the SLy6 interaction as shown in Fig.~\ref{fig:EoS-protfrac}. 
Pure neutron matter ($X_p=0$) is unbound, a feature which is well known. Unbound 
matter persists for small $X_p$ up to about $X_p=0.09$. For $X_p=0.1$ we are 
able to find a local minimum at non-zero density which, however, is a metastable 
state. A well bound ground state (negative energy) emerges then from $X_p=0.13$ 
on. Binding and equilibrium densities increase very quickly above $X_p=0.13$. We 
thus have a clear distinction between low proton content $X_p$ and moderate or 
large one with a proton fraction around a critical point $X_p\approx 0.13$. Once 
substantial binding sets on above the critical point, the equilibrium densities 
saturate quickly around $\rho_\mathrm{nm}=0.12-0.16$ fm$^{-3}$. Just recently a 
study appeared of the equation-of-state of asymmetric nuclear matter based on 
classical molecular dynamics simulations \cite{Lop14a}. Although the formal 
framework is very different, the classical results at temperatures as low as 2 
MeV show the same trends as seen here at zero temperature, e.g., the shape of 
the binding curves and the drift of the equilibrium density with $X_p$.

The sequence of equations of state for different $X_p$ looks very similar for 
all Skyrme interactions which we have studied. It can be characterized by the 
sequence of equilibrium points shown as black boxes in 
Fig.~\ref{fig:EoS-protfrac}. Fig.~\ref{fig:vary-asymnucmat} shows this sequence 
of equilibrium points for a great variety of Skyrme interactions. The upper 
panel shows results for widely used interactions found in the literature, SkM* 
\cite{Bar82a}, SLy6 \cite{Chabanat}, SkI3 \cite{Rei95a}, and UNEDF0 
\cite{Kor10}. The lower panel shows results from a set of interactions in which 
nuclear matter properties have been systematically varied with respect to 
standard values set in SV-bas \cite{Klu09a}, SV-mas07 with lower effective mass, 
SV-K218 with smaller incompressibility, SV-sym34 with larger symmetry energy, 
and SV-kap00 with smaller Thomas-Reiche-Kuhn sum rule enhancement factor 
(equivalent to isovector effective mass). All of the interactions show very 
similar behavior. The largest deviations are seen for the interactions with 
untypically large symmetry energy, SkI3 in the upper panel and SV-sym34 in the 
lower one. Even including these, all interactions show the same trends and the 
binding energies in dependence on the proton fraction are almost identical. 
Therefore we expect the results in this work obtained with SLy6 representative 
for all reasonable Skyrme interactions. Varying the interaction may shift 
slightly the borders between phases. But the overall sequence should be robust.

\subsection{The map of pasta shapes}

The map of pasta shapes obtained from the dynamical simulations under various
initial conditions is shown in the upper panel of Fig.~\ref{fig:phadiag_X_P}. 
For every value of $X_p$ and mean density $\rho$, two calculations with
different initial conditions are performed. These yield in most cases the same
phase. There are few cases where two different final states are reached. This is
indicated by showing two different colors in a cell.  The value $X_p=0.3$ is
replaced by $X_p=1/3$ to establish contact with the previous study
\cite{Schuetrumpf2013a} which worked exclusively at this proton fraction. To
fill the resulting larger gap towards $X_p=0.2$, we also show a lower value
$X_p=0.29$.  At first glance, one sees a jump between $X_p=0.1$ and the larger
$X_p$ to the extent that $X_p=0.1$ has a much larger range of pure nuclear
matter on the side of high $\rho$.  This reflects the ``phase transition''
around $X_p\approx 0.13$ observed in the equation of state in
Fig.~\ref{fig:EoS-protfrac}. The transition between pure nuclear matter and
structured matter is, in fact, predominantly a competition between the spherical
bubbles (black boxes) and homogeneous matter. All other shapes show steady,
often moderate, changes with $X_p$. The sequence in which the shapes appear and
disappear with increasing density is practically the same at all $X_p$. There is
a slight trend, however, that for an increasing proton fraction the different
pasta shapes appear at lower mean densities. Take, e.g. the border between rod
and rod(2): it is above a mean density of $\rho=0.05{\rm\,fm^{-3}}$ for
$X_p=0.1$ and moves to about $0.04{\rm\,fm^{-3}}$ for symmetric matter
($X_p=0.5$).

In Ref.~\cite{Okamoto} the trend that for low proton fractions pasta matter
appears in a smaller range in density is also clearly visible. The densities for
the transitions between normal nuclei, pasta matter, and uniform nuclear matter
are in detail different, like in other approaches. These numbers seem to be very
sensitive to the equation of state used for the calculations.

It is interesting to compare the present results at $X_p=0.33$ to our earlier
simulations \cite{Schuetrumpf2013a}. The sequence of shapes is the same. But the
middle region shows more slab configurations. This is probably an effect of
different initialization.  The earlier calculation allowed a smaller minimal
distance between the $\alpha$-particles leading to an initial state which
contains more clustered $\alpha$-particles obviously driving the slab phase. The
appearance of shapes thus seems to depend to some extent on the initial state.
This indicates that the barriers between the different shapes are sufficiently
high to stabilize a shape even though it is not necessarily the minimum
configuration.

As a further check of the stability of the dynamical configurations, we have
cooled each of them down to temperature zero actually starting the static
iteration from the final stage of dynamical evolution. The resulting map of
shapes is shown in the lower panel of Fig.~\ref{fig:phadiag_X_P}. Most of the
dynamical configurations persist when cooled down. There are only a few small
shifts of borders between shapes and, not surprisingly, a few more slabs pop up
which corrects somewhat the underestimation of slabs. 

A subtle difference in detail ought to be mentioned.  The border to uniform
nuclear matter is shifted to smaller mean densities for $X_p=0.1$ but to higher
mean densities for all higher proton fractions $X_p>0.1$. The jump in the border
between uniform nuclear matter and spherical bubbles thus becomes even more
pronounced. In fact, the larger jump for the cooled configurations is closer to
the sharp transition indicated in the equation of state, see
Fig.~\ref{fig:EoS-protfrac}.  Increasing temperature weakens this transition.

\subsection{Neutron background}

\begin{figure}[tb]
 \centering
\includegraphics[width=\columnwidth]{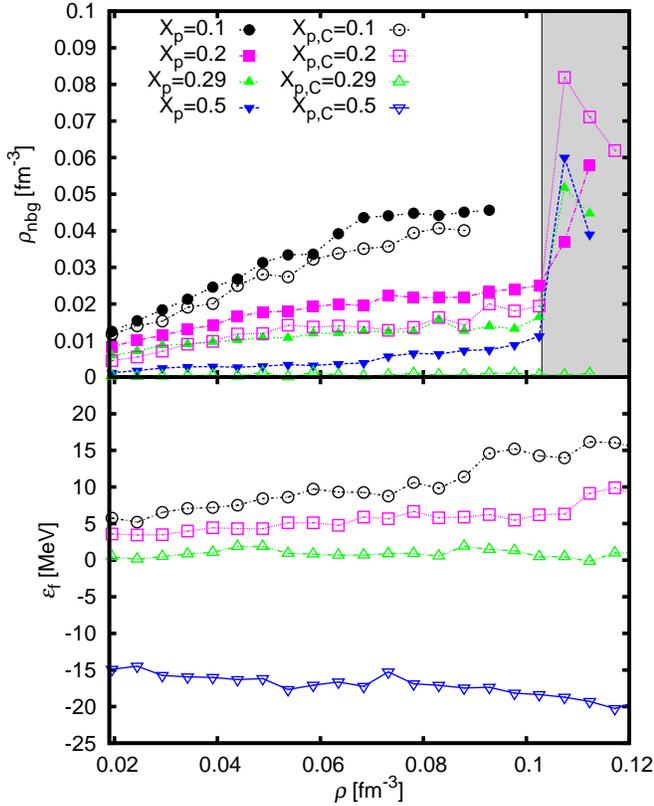}
\caption{\label{fig:X_P_nbg}(Color online)  Upper panel: Neutron background densities
($\rho_{\rm nbg}$) of the series of calculations of Fig.~\ref{fig:phadiag_X_P}.
Full symbols show results from dynamical calculations and open symbols from
zero-temperature configurations, marked with the additional index C in the
legend. The case $X_{p,C}=0.5$ is not shown because it did not produce a
background. The shaded area indicates the region where $\rho_g$ is only vaguely
defined. Lower panel: Neutron Fermi energies for the cooled Pasta shapes in
dependence on the mean density for different proton fractions are shown.}
\end{figure}

We have matter with very different neutron contents. It is thus interesting to
look for the gaseous neutron background density $\rho_g$ under the different
conditions.  In order to find $\rho_g$, we follow the strategy of
Ref.~\cite{Schuetrumpf2013a} and compute the distribution of volumes of the
neutron density, $v(\rho_\mathrm{ref})=\int d^3r\, \delta(\rho_\mathrm{ref}
-\rho_n(\mathbf{r}))$.  This displays typically a clear peak at low
$\rho_\mathrm{ref}$.  A Gauss fit to this low density peak in the neutron
distribution then yields $\rho_g$. No clear peak is visible for higher mean
densities.  In this case, the edge where the curves starts to differ from zero
was taken as an indicator. The proton background densities, if present at all,
are very small and can be ignored. The results are shown in the upper panel of
Fig.~\ref{fig:X_P_nbg}. The shaded area at high $\rho$ indicates the region
where the determination of $\rho_g$ becomes uncertain. Naturally, $\rho_g$ is
very large for the smallest proton fraction $X_p=0.1$ and drops quickly with
increasing $X_p$. The values from cooled configurations are systematically lower
than for the dynamical ones. This reflects the fact that cold bound systems can
accommodate more extra neutrons than hot ones. For $\rho\gtrsim
0.1{\rm\,fm^{-3}}$ (shaded area) the tails of the density distributions in the
voids of the rod bubble or the sphere bubble cover very little space and thus do
not reach small values for the density. Therefore in these configurations a
large neutron background is present.

While the neutron background is present for all proton fractions in the finite
temperature calculations, it vanishes completely for $X_p>0.29$ for the zero
temperature calculations. In order to corroborate this result, we show in the
lower panel of Fig.~\ref{fig:X_P_nbg} the neutron Fermi energies
$\epsilon_{F,n}$ for the cooled configurations. For $X_p=0.1$ and $X_p=0.2$, the
$\epsilon_{F,n}$ stay significantly above zero, confirming that a large amount
of neutrons are unbound. The case of $X_p=0.29$ is transitional to the extent
that the $\epsilon_{F,n}$ are close to zero and therefore, the neutron
background is only marginally present. For $X_p=0.33$ the Fermi energies are
slightly below zero and thus there is no neutron background present any more.
For even higher $X_p$ the Fermi energies drop further below zero and yield a
strongly bound system. For finite temperature calculations the distribution
around the Fermi edge is softened and therefore a small neutron background is
present even for large proton fractions.

\subsection{Liquid phase}

\begin{figure}[tb]
 \centering
\includegraphics[width=\linewidth]{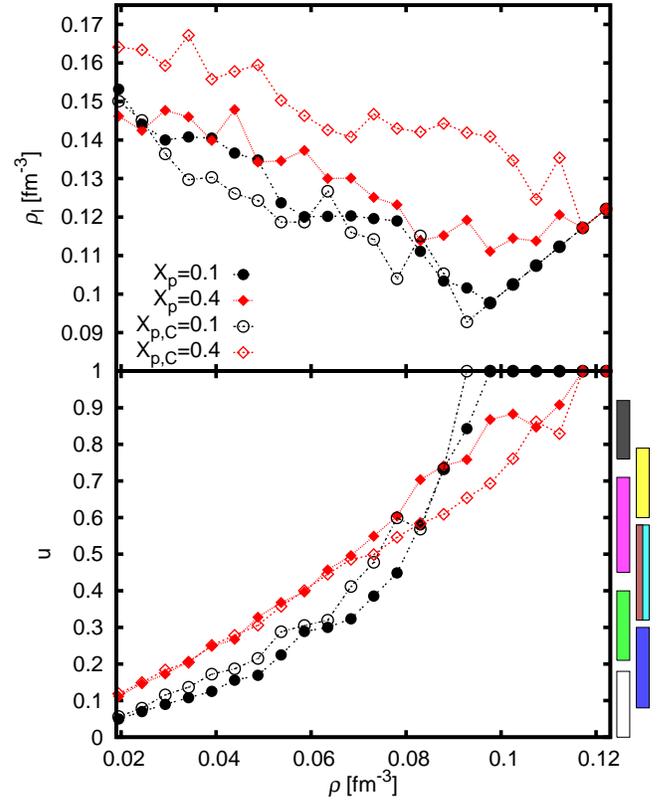}
\caption{\label{fig:rho_l}(Color online) Upper panel: Liquid phase densities for two
  representative values of $X_p$.  The cooled calculations are marked
  with the additional index C.  Lower panel: Liquid phase occupied
  volume fractions $u_l$ for two $X_p$.  The color bars on the right
  indicate a pasta shape with the color code as explained in
  Fig.~\ref{fig:phadiag_X_P}.  The bars mark the intervals of $u_l$
  for which the pasta shape associated with this color appears.}
\end{figure}
The liquid densities $\rho_l$ are also computed from the distribution of the
total density $v(\rho_\mathrm{ref})$, but now fitting the high density peaks or
flanks.  The results for $\rho_l$ are shown in the upper panel of
Fig.~\ref{fig:rho_l}. Only graphs for two values of $X_p$ are shown because
there is very little variation of the $\rho_l$ for $X_p=0.2...0.5$.  The line
for $X_p=0.4$ is taken to represent all lines for $X_p\geq0.2$. Visibly
different is the case $X_p=0.1$ where $\rho_l$ is lower than for the other $X_p$
for both cases, cooled and dynamic. This complies well with the equilibrium
densities shown in Fig.~\ref{fig:EoS-protfrac} which also shows this marked
difference between low $X_p\leq0.1$ and larger $X_p>0.13$. The actual values for
$\rho_l$ can differ from the case of homogeneous matter, because the liquid
phase is not as neutron rich as the proton fraction indicates due to the
separate neutron background and due to surface and Coulomb effects. The liquid
phase density decreases with increasing mean density and converges to the mean
density below nuclear saturation density.  Not surprisingly, the cooled
calculations show higher liquid densities because they are better bound. The
effect is much more pronounced for high $X_p$.

From $\rho_l$ and $\rho_g\approx\rho_{\rm nbg}$ the volume fraction $u_l$ can be
derived using Eq.~(\ref{eq:Gibbs}). The result is displayed in the lower panel
of Fig.~\ref{fig:rho_l}.  Again, we take one representative $X_p=0.4$ for the
whole group $X_p\geq 0.2$. Most curves show an increasing slope, so they grow
faster than linearly. This correlates to the fact that the liquid density
decreases with increasing density (upper panel of Fig.~\ref{fig:rho_l}) which
results in a larger volume fraction. Very interesting is the behavior of the
curves for $X_p=0.1$. There are two counteracting effects. First, we have a low
liquid density which results in a larger volume fraction. Second, $\rho_g$ is
very large which reduces the volume fraction. The figure shows that for low mean
densities the background neutron effect dominates, resulting in lower volume
fraction than for the other $X_p$. For higher mean densities the small values
for $\rho_l$ dominate and the $u_l$ for $X_p=0.1$ crosses the other lines making
the transition from low $u_l$ to high ones significantly steeper than for the
larger $X_p$.  This can also be seen in Fig.~\ref{fig:phadiag_X_P}: The sequence
of non-trivial shapes (rod until spherical-bubble) is compressed to a
smaller range of $\rho$ than for $X_p\geq 0.2$.

The impact of the type of the Skyrme force on these results can be estimated
from Fig.~\ref{fig:vary-asymnucmat}. The main difference is that the liquid
densities would be different especially for SkI3 and $a_{\rm sym}=34{\rm\,MeV}$.
Therefore the values in Eq.~\ref{eq:Gibbs} change slightly and the transition
points from one to another pasta structure would change. The overall structure
of the map of pasta should be the same for all interactions.

\section{Conclusion}

Using time-dependent Hartree-Fock simulations and static Hartree-Fock
calculations, we have investigated the scenarios of the various ``pasta''
configurations in nuclear matter under astrophysical conditions. Thereby, we
have explored the appearance of the different shapes (sphere, rod, slab, ...,
bubble) in dependence on a given mean density  and proton fraction. The
dynamical simulations produce thermal states with temperatures in the range 2--7
MeV. The static calculations start from the thermal states and serve to check
the stability of geometry under changing thermal conditions.

We find a clear distinction between a regime of low proton fractions $X_p\leq
0.13$ and high ones. The results for $X_p=0.2...0.5$ show practically the same
shapes as function of mean density $\rho$ while the sequence of non-trivial
geometries is compressed to a smaller density range for $X_p=0.1$. The sequence
as such is similar in all cases, starting from isolated spheres (nuclei) at very
low density, proceeding over rods, planar meshes of orthogonal rods, triaxial
meshes of rods, slabs, to the reciprocal profiles of planar meshes of rod-like
bubbles, linear bubble-rods, spherical bubbles, and finally homogeneous matter.
The special properties of the small low-density regime are corroborated from the
equation of state of homogeneous matter. Matter is unbound for low $X_p$.
Binding sets on at around $X_p=0.13$ and develops very quickly to strong binding
with an almost constant equilibrium density.

Generally, the matter segregates into high-density regions of nuclear quantum
liquid and low-density regions of a gas of background nucleons, predominantly
neutrons. Its density increases with mean density $\rho$ and decreases with
increasing $X_p$. There is also a significant differences between thermal and
cooled state to the extent that the cooled state contains much less neutron gas,
for $X_p>0.29$ even no gas at all. These observations are corroborated by the
neutron Fermi energies which are above zero in the cases of largest neutron gas
background.

The volume fraction $u_l$ (liquid volume divided by total volume) is the most
important parameter determining the pasta shapes. Each shape is found to have a
certain interval of $u_l$ for which it can appear. The trend $u_l(\rho)$ shows
also a clear distinction between $X_p=0.1$ and all larger $X_p$. The curve is
steeper for the low $X_p$ which is related to the compressed sequence of shapes
at $X_p=0.1$. 

All in all, the results are rather robust in the region of larger proton
fraction $X_p=0.2...0.5$. This means earlier investigations done for one proton
fraction, mostly $X_p=1/3$ apply in this whole range. The region of low proton
fractions is different and generally more sensitive to the system parameters.

\smallskip

\begin{acknowledgments}
This work was supported by the Bundesministerium f\"ur Bildung und Forschung
(BMBF) under contract number 05P12RFFTB and by Grants-in-Aid for Scientific
Research on Innovative Areas through No. 24105008 provided by MEXT.
\end{acknowledgments}

\bibliography{ProtonFraction}

\end{document}